\documentclass[11pt,a4paper]{article}

\usepackage[T1]{fontenc}
\usepackage[utf8]{inputenc}
\usepackage{amsmath,amssymb,amsfonts,amsthm}
\usepackage{graphicx}
\usepackage{booktabs}
\usepackage{array,tabularx}
\usepackage{placeins}
\usepackage{float}
\usepackage[margin=2.6cm]{geometry}
\usepackage[authoryear,round]{natbib}
\usepackage[hidelinks]{hyperref}

\theoremstyle{plain}
\newtheorem{theorem}{Theorem}[section]
\newtheorem{proposition}[theorem]{Proposition}
\newtheorem{lemma}[theorem]{Lemma}
\newtheorem{corollary}[theorem]{Corollary}
\theoremstyle{definition}
\newtheorem{definition}[theorem]{Definition}
\newtheorem{assumption}[theorem]{Assumption}
\newtheorem{example}[theorem]{Example}
\theoremstyle{remark}
\newtheorem{remark}[theorem]{Remark}

\newcommand{\B}{\mathcal{B}}
\newcommand{\R}{\mathcal{R}}
\newcommand{\Tk}[1]{\mathfrak{T}_{#1}}
\newcommand{\RB}{\mathfrak{R}_{\B}}
\newcommand{\CB}{\mathfrak{C}_{\B}}
\newcommand{\RR}{\mathfrak{R}_{\R}}
\newcommand{\CR}{\mathfrak{C}_{\R}}
\newcommand{\Dvec}{\langle D_{\mathfrak{T}}\rangle}
\newcommand{\Avec}{\langle A_{\mathfrak{T}}\rangle}
\newcommand{\simp}{\mathcal{A}}
\newcommand{\Nset}{\mathcal{N}}

\begin{document}

\title{\bfseries Continuation-Adjusted Full-Support Indifference in a
Feedback Stackelberg Attacker--Defender Attrition Game}

\author{Azhar Iqbal\thanks{Corresponding author: \texttt{azhar.iqbal@adelaide.edu.au}.}
\and James M.\ Chappell \and Derek Abbott}

\date{School of Electrical and Mechanical Engineering,\\
Adelaide University, Adelaide, South Australia 5005, Australia}

\maketitle

\begin{abstract}
Successful attacks that remove assets break two familiar static regularities:
coverage need not deter attack, and the defender need not push coverage until
its budget or an individual target saturates.
We analyse a finite-horizon attacker--defender attrition game under a stagewise
feedback-recommitment strong Stackelberg equilibrium. Our central result is a
continuation adjustment that converts each state--period node into a static
security game with shifted payoffs. At feasible full-support nodes, coverage
remains indexed by one scalar attacker value, so backward induction uses scalar
interval and endpoint calculations; boundary-support nodes require conditional
linear programmes. Continuation losses can reverse an attacker's coverage
swing, making a covered asset more attractive because it survives for later
attacks. A three-asset, two-period instance is solved globally for every discount
factor. It exhibits an exact singular threshold at which the optimal set
expands, the selected policy jumps, and the defender abandons a previously
protected asset and leaves budget unused with no target saturated, although
both players' values remain continuous.
\end{abstract}



\section{Introduction}
\label{sec:intro}

Stackelberg security games model a defender who commits to a protective
allocation before an observing attacker chooses a target
\citep{ConitzerSandholm2006,Kiekintveld2009,Korzhyk2011,Tambe2011}. In the
standard static model, covering a target lowers its value to the attacker, and
with positive payoff swings the defender uses the available coverage budget.
These properties need not survive when a successful attack removes an asset and
thereby changes the future target set.

We study a finite-horizon attrition game whose state is the set of surviving
assets. At each state--period node the defender recommits, the attacker observes
the allocation, and ties are resolved by strong Stackelberg selection with an
explicit lexicographic refinement. The paper's principal contribution is a
continuation-adjustment lemma: after adding state-contingent continuation values
to the primitive payoffs, each dynamic node has exactly the affine target-payoff
form of a static security game. Consequently, whenever the attacker's
full-support indifference locus is feasible, every coverage coordinate is
indexed by the single common attacker value. Feasibility becomes a scalar
interval calculation and strong tie-breaking reduces selection on that interval
to an endpoint comparison. Boundary-support nodes remain ordinary
support-specific linear programmes.

The adjustment reveals a specifically dynamic mechanism. Destroying an asset
may sacrifice profitable future attacks on it. When that continuation loss
exceeds the immediate benefit of destruction, the attacker's coverage swing
changes sign: covering the asset makes an immediate strike more attractive
because the asset survives. This separates deterrence from protection and can
make unused defensive capacity optimal even when all primitive payoff swings
are positive.

A three-asset, two-period instance makes these effects exact. We derive the
global strong Stackelberg equilibrium for every discount factor, solve the
singular node at which one adjusted swing vanishes, and characterise the full
optimal set there. At the threshold, both players' values remain continuous but
the selected policy jumps: the defender drops coverage of the asset protected
most heavily by the myopic policy and begins to leave budget unused. The
optimal-set correspondence is upper but not lower hemicontinuous at that point.

The static full-support formulae used by the reduction are established security-
game machinery and are presented only in the form needed for the dynamic
analysis. The method is not a general closed-form solver for dynamic
Stackelberg games: it applies directly at qualifying full-support nodes, while
boundary supports require conditional linear programmes and the subset state
space remains exponential. Sections~\ref{sec:model}--\ref{sec:budget} develop
the compact node solver; Section~\ref{sec:dynamic} contains the continuation
adjustment, sign reversal and exact threshold analysis.

\section{Related work}
\label{sec:related}

Classical Stackelberg security games allocate scarce protective resources after
a defender commitment and before an observed attacker response
\citep{Kiekintveld2009,Korzhyk2011,Tambe2011,Kar2018,Sinha2018}, and have been
fielded for airport patrols, air marshals, port security and wildlife
protection \citep{Pita2008,Jain2010,Shieh2012,Yang2014}. Defence-oriented
Stackelberg formulations include procurement models \citep{Hohzaki2009}, and
general treatments of dynamic games and their solution concepts are collected in
\citet{BasarZaccour}. Strong
Stackelberg equilibria can be computed by solving one target-conditional linear
programme for each possible attacked target \citep{ConitzerSandholm2006}. The
full-support formula used below is the familiar common-attacker-payoff
parameterisation underlying specialised security-game algorithms such as
\textsc{Origami} \citep{Kiekintveld2009}. We use these static results as a node
solver, not as a new static solution theory.

Dynamic and stochastic security models commonly require value iteration,
policy iteration, learning, or nested optimisation. Examples include Markov
moving-target defence \citep{LiShenZheng2020}, dynamic attacker engagement
\citep{Pawlick2019}, repeated stochastic asymmetric interaction
\citep{Cheng2023}, approximate feedback Stackelberg computation in nonlinear
games \citep{LiSojoudi2024}, policy iteration for stochastic Stackelberg games
\citep{Kudo2025}, and no-regret learning in dynamic Stackelberg games
\citep{Lauffer2024}. Existence and uniqueness questions have also been studied
for feedback Stackelberg equilibria on controlled Markov processes
\citep{BressanJiang2020} and for discounted stochastic Stackelberg games
\citep{WuZhang2026}.

Our setting is narrower: discrete time, finite horizon, complete information,
and a state that only shrinks as assets are destroyed. It also uses stagewise
feedback recommitment rather than a single initial commitment to a contingent
policy. This restriction permits an exact continuation adjustment and exposes a
mechanism not visible in the static primitives: continuation values can reverse
the sign of an attacker's coverage swing. The resulting closed form applies at
feasible full-support nodes, while general boundary-support nodes retain the
standard conditional-LP calculation. Related attacker--defender and
interdiction models are usually formulated as bilevel or trilevel optimisation
problems \citep{Brown2006,SmithLim2008,SmithSong2020,Hunt2024}; the present
contribution is instead an analytical benchmark that isolates how attrition
changes the local Stackelberg geometry. It extends the special backward-induction setting
of \citet{Iqbal2024} by treating arbitrary finite target sets, exact marginal-
coverage feasibility, set-valued follower responses, and the singular dynamic
threshold.

\section{The model}
\label{sec:model}

A defender $\B$ protects $N$ assets indexed by
$\Nset=\{1,\ldots,N\}$ against an attacker $\R$. In the unit-allocation model,
the defender chooses $D\in\simp_N$ and the attacker chooses $A\in\simp_N$, where
\begin{equation}
\simp_N=\Big\{x\in\mathbb{R}_{+}^{N}:\ \sum_{n=1}^{N}x_n=1\Big\}.
\label{eq:simplex}
\end{equation}
Here $D_n$ is the marginal probability that target $\Tk{n}$ is covered and
$A_n$ is the probability that it is attacked.

With $M$ identical resources, no routing or scheduling constraints, and idling
allowed, the implementable marginal-coverage set is
$\{D\in[0,1]^N:\sum_nD_n\le M\}$, the convex hull of incidence vectors of target
subsets of cardinality at most $M$. This is a coverage probability, not an
assignment share averaged across resources. If all resources must be deployed,
the inequality is replaced by equality. Richer operational constraints would
produce a smaller marginal polytope \citep{Korzhyk2010,Korzhyk2011}.

\subsection{Payoffs}
\label{sec:payoffs}

Let $\RB(\Tk{n})$ be the reward to $\B$ if the attacked tank $\Tk{n}$ is
protected, $\CB(\Tk{n})$ the loss parameter for $\B$ if it is unprotected,
$\RR(\Tk{n})$ the reward to $\R$ if it is unprotected, and $\CR(\Tk{n})$ the
loss parameter for $\R$ if it is protected. All are common knowledge. If
$\Tk{n}$ is attacked, the payoff to $\B$ is
$D_{n}\RB(\Tk{n})-(1-D_{n})\CB(\Tk{n})$ and that to $\R$ is
$(1-D_{n})\RR(\Tk{n})-D_{n}\CR(\Tk{n})$. Weighting by the attack probabilities,
\begin{align}
\Pi_{\B}\{\Dvec,\Avec\}
&=\sum_{n=1}^{N}A_{n}\big\{D_{n}\RB(\Tk{n})-(1-D_{n})\CB(\Tk{n})\big\},
\label{eq:PiB}\\
\Pi_{\R}\{\Dvec,\Avec\}
&=\sum_{n=1}^{N}A_{n}\big\{(1-D_{n})\RR(\Tk{n})-D_{n}\CR(\Tk{n})\big\}.
\label{eq:PiR}
\end{align}
A tank attacked with probability zero contributes nothing to either payoff. The
model abstracts from attacker operating costs.

Write the payoff swings
\begin{equation}
\Omega_{n}^{\R}=\RR(\Tk{n})+\CR(\Tk{n}),\qquad
\Omega_{n}^{\B}=\RB(\Tk{n})+\CB(\Tk{n}),
\label{eq:omegas}
\end{equation}
which measure how far each player's payoff at target $n$ moves as that target
goes from unprotected to protected. The per-target payoffs are then affine in
$D_{n}$:
\begin{equation}
u_{n}(D):=\RR(\Tk{n})-D_{n}\Omega_{n}^{\R},\qquad
g_{n}(D):=D_{n}\Omega_{n}^{\B}-\CB(\Tk{n}),
\label{eq:un_gn}
\end{equation}
so $\Pi_{\R}=\sum_{n}A_{n}u_{n}(D)$ and $\Pi_{\B}=\sum_{n}A_{n}g_{n}(D)$. Thus
$\Omega_{n}^{\R}>0$ means protection strictly reduces the attacker's payoff at
$\Tk{n}$, while $\Omega_{n}^{\R}<0$ means it strictly increases it.

\begin{remark}[Signs]
\label{rem:signs}
The symbol $\mathfrak{C}$ is retained for consistency with the cost terminology
of \citep{Iqbal2024}, but these are signed payoff parameters. We call an instance
\emph{sign-regular} when $\Omega_{n}^{\R}>0$ for every $n$, which holds in
particular whenever $\RR,\CR\ge0$ and are not both zero at any target. By
Proposition~\ref{prop:signdichotomy} the degenerate case of
Proposition~\ref{prop:feasibility}\,(iii) occurs only outside that regime.
\end{remark}

\begin{assumption}
\label{ass:nonzero}
$\Omega_{n}^{\R}\neq0$ for every $n=1,\ldots,N$.
\end{assumption}

If $\Omega_{n}^{\R}=0$ then $u_{n}$ does not depend on $D_{n}$, so the defender
has no leverage over the attacker's payoff at target $n$ and that target can
belong to a full-support configuration only for the isolated parameter value
$\RR(\Tk{n})=v$ of Lemma~\ref{lem:symmetric}.

\section{The two-stage game and solution concept}
\label{sec:twostage}

At stage~1, $\B$ chooses $\Dvec\in\simp_{N}$; at stage~2, $\R$ observes that
vector and chooses $\Avec\in\simp_{N}$; terminal payoffs are \eqref{eq:PiB} and
\eqref{eq:PiR}. This is a finite-horizon dynamic game of complete and perfect
information \citep{BasarOlsder,FudenbergTirole,Osborne}.

Since $\Pi_{\R}$ is linear in $A$, the follower's maximiser need not be unique,
and we work with the best-response correspondence
\begin{equation}
BR_{\R}(\Dvec)=\arg\max_{\Avec\in\simp_{N}}\Pi_{\R}\{\Dvec,\Avec\}
=\operatorname{conv}\big\{e_{n}:\ n\in\textstyle\arg\max_{1\le k\le N}u_{k}(D)
\big\},
\label{eq:BR}
\end{equation}
the face of $\simp_{N}$ spanned by the unit vectors $e_{n}$ of the targets of
maximal per-target attacker payoff.

We adopt the strong Stackelberg convention standard in security games
\citep{ConitzerSandholm2006,Korzhyk2011,Tambe2011}: ties in the follower's
best-response set are broken in the leader's favour.  For a defender choice
$D$, define
\begin{equation}
q(D)=\max_n u_n(D),\qquad
W(D)=\max\{g_n(D):u_n(D)=q(D)\}.
\label{eq:qW}
\end{equation}
Thus $q(D)$ is the follower value and $W(D)$ is the defender payoff after strong
tie-breaking.  A strong Stackelberg defender choice maximises $W$ over the
feasible coverage set $\mathcal D$.

For the dynamic recursion we use the following refinement, which does not alter
the SSE defender value.  Among all defender choices maximising $W$, select one
that minimises $q(D)$; among any remaining value-equivalent choices, report a
resource-minimal representative, and finally use a fixed lexicographic order on
$D$ if needed.  Equivalently, the selected outcome lexicographically maximises
\begin{equation}
\big(W(D),-q(D),-\textstyle\sum_nD_n,-D_1,\ldots,-D_N\big).
\label{eq:lexSSE}
\end{equation}
The first component is the usual optimistic/strong Stackelberg criterion.  The
secondary components are an explicit equilibrium-selection convention: they
make both players' continuation values and the displayed policy single-valued
without asserting that all SSE commitments are outcome-equivalent.  In the
unit-budget model the resource-minimal component is vacuous.

An SSE outcome is therefore a pair $(D^*,A^*)$ with
$A^*\in BR_{\R}(D^*)$, chosen to attain $W(D^*)$, and $D^*$ selected by
\eqref{eq:lexSSE}.  Under a weak convention, or another secondary selection
rule, the full-support feasibility analysis is unchanged, although dynamic
continuation values and the reported representative policy can differ at
value-degenerate nodes.

\begin{remark}[Existence]
\label{rem:existence}
For each feasible $D$, $BR_{\R}(D)$ is a nonempty compact face of the attack
simplex and the strong-tie-break value is
$W(D)=\max\{g_n(D):n\in\arg\max_k u_k(D)\}$. Because the target set is finite,
$W$ is upper semicontinuous: along any convergent sequence one may pass to a
subsequence with a fixed selected target, whose best-response inequalities and
payoff survive in the limit. Hence $W$ attains its maximum on the compact
coverage set. The continuous secondary criteria in \eqref{eq:lexSSE}, applied
successively on nested compact optimizer sets, select a unique coverage vector.
The same argument applies at every dynamic node.
\end{remark}

\section{The full-support indifference locus}
\label{sec:indifference}

Since $\Pi_{\R}=\sum_{n}A_{n}u_{n}(D)$ is linear in $A$, the attacker places
positive probability only on targets maximising $u_{n}(D)$. Hence an attack
vector with full support is a best response if and only if
\begin{equation}
u_{1}(D)=u_{2}(D)=\cdots=u_{N}(D).
\label{eq:indiff}
\end{equation}
This is a support condition for a linear programme, not a first-order condition
for an interior maximum: if \eqref{eq:indiff} fails, the best response
concentrates on the targets of largest per-target payoff.

\begin{definition}[Full-support indifference locus]
\label{def:locus}
The \emph{full-support indifference locus} is the affine set
\begin{equation}
\mathcal{L}=\big\{D\in\mathbb{R}^{N}:\ u_{1}(D)=u_{2}(D)=\cdots=u_{N}(D)\big\},
\label{eq:locus}
\end{equation}
and its \emph{feasible part} is $\mathcal{L}\cap\mathcal{D}$, where $\mathcal{D}$
is the defender's feasible set: $\simp_{N}$ in
Sections~\ref{sec:indifference}--\ref{sec:envelope} and $\mathcal{D}_{N,M}$ in
Section~\ref{sec:budget}. The qualifier ``full-support'' refers to the
\emph{attacker's} strategy, not the defender's: by \eqref{eq:BR},
$D\in\mathcal{L}$ if and only if $BR_{\R}(D)=\simp_{N}$, so that every attack
vector, in particular every one with full support, is a best response. It does
not require $D$ itself to have full support, and in general the selected $D$
does not: the vector selected in Example~\ref{ex:four} has $D_{1}=0$.
\end{definition} Eliminating
$A_{N}=1-\sum_{n<N}A_{n}$ gives the equivalent familiar form
$\partial\Pi_{\R}/\partial A_{n}=u_{n}(D)-u_{N}(D)=0$, $1\le n\le N-1$.

Taking $\Tk{N}$ as reference in \eqref{eq:indiff} and using \eqref{eq:un_gn},
Assumption~\ref{ass:nonzero} gives
\begin{equation}
D_{n}=\frac{\RR(\Tk{n})-\RR(\Tk{N})+\Omega_{N}^{\R}D_{N}}{\Omega_{n}^{\R}},
\qquad 1\le n\le N-1,
\label{eq:Ds}
\end{equation}
so that with
\begin{equation}
a_{n}:=\frac{\RR(\Tk{n})-\RR(\Tk{N})}{\Omega_{n}^{\R}},\qquad
b_{n}:=\frac{\Omega_{N}^{\R}}{\Omega_{n}^{\R}},\qquad 1\le n\le N-1,
\label{eq:anbn}
\end{equation}
the locus is the affine family
\begin{equation}
D_{n}=a_{n}+b_{n}D_{N},\qquad 1\le n\le N-1.
\label{eq:affine}
\end{equation}
The parametrisation \eqref{eq:affine} singles out a target that plays no
distinguished role. The following reformulation removes it, and it is the form
the dynamic reduction of Section~\ref{sec:dynamic} uses.

\begin{lemma}[Reference-free description]
\label{lem:symmetric}
Let Assumption~\ref{ass:nonzero} hold, let $v\in\mathbb{R}$, and set
\begin{equation}
\sigma:=\sum_{n=1}^{N}\frac{1}{\Omega_{n}^{\R}},\qquad
\rho:=\sum_{n=1}^{N}\frac{\RR(\Tk{n})}{\Omega_{n}^{\R}}.
\label{eq:sigmarho}
\end{equation}
Then $D\in\mathbb{R}^{N}$ lies in $\mathcal{L}$ with common value
$u_{1}(D)=\cdots=u_{N}(D)=v$ if and only if
\begin{equation}
D_{n}=\frac{\RR(\Tk{n})-v}{\Omega_{n}^{\R}},\qquad n=1,\ldots,N,
\label{eq:Dv}
\end{equation}
in which case
\begin{equation}
\sum_{n=1}^{N}D_{n}=\rho-\sigma v,
\qquad\text{so that}\qquad
\sum_{n=1}^{N}D_{n}=1
\ \Longleftrightarrow\
\sigma v=\rho-1.
\label{eq:scalar}
\end{equation}
Moreover, for any choice of reference index $N$ in \eqref{eq:anbn},
\begin{equation}
1+\sum_{n=1}^{N-1}b_{n}=\Omega_{N}^{\R}\sigma,\qquad
1-\sum_{n=1}^{N-1}a_{n}=1-\rho+\RR(\Tk{N})\sigma .
\label{eq:identities}
\end{equation}
\end{lemma}

\begin{proof}
By \eqref{eq:un_gn} and Assumption~\ref{ass:nonzero}, $u_{n}(D)=v$ for all $n$
is equivalent to $\RR(\Tk{n})-D_{n}\Omega_{n}^{\R}=v$ for all $n$, that is to
\eqref{eq:Dv}. Summing \eqref{eq:Dv} over $n$ gives
$\sum_{n}D_{n}=\rho-v\sigma$, whence \eqref{eq:scalar}. For
the first identity,
$1+\sum_{n<N}\Omega_{N}^{\R}/\Omega_{n}^{\R}
=\Omega_{N}^{\R}\big(1/\Omega_{N}^{\R}+\sum_{n<N}1/\Omega_{n}^{\R}\big)
=\Omega_{N}^{\R}\sigma$. For the second,
$\sum_{n<N}a_{n}
=\big(\rho-\RR(\Tk{N})/\Omega_{N}^{\R}\big)
-\RR(\Tk{N})\big(\sigma-1/\Omega_{N}^{\R}\big)=\rho-\RR(\Tk{N})\sigma$.
\end{proof}

Equations \eqref{eq:Dv}--\eqref{eq:scalar} involve no distinguished target, so
the solution set does not depend on the reference index even though
$a_{n},b_{n}$ do; and $v$ is the attacker's equilibrium payoff on the locus,
since $D$ satisfying \eqref{eq:Dv} gives $\Pi_{\R}\{D,A\}=v$ for every
$A\in\simp_{N}$.

\section{Feasibility of the protection vector}
\label{sec:feasibility}

The algebra of Section~\ref{sec:indifference} produces a locus of solutions to
the indifference equations; it does not guarantee that any point of that locus
is a probability vector. From \eqref{eq:affine}, normalisation reads
\begin{equation}
\Big(1+\sum_{n=1}^{N-1}b_{n}\Big)D_{N}=1-\sum_{n=1}^{N-1}a_{n},
\label{eq:norm}
\end{equation}
which, when $1+\sum_{n<N}b_{n}\ne0$, determines
\begin{equation}
D_{N}^{*}=\frac{1-\sum_{n=1}^{N-1}a_{n}}{1+\sum_{n=1}^{N-1}b_{n}}.
\label{eq:DNstar}
\end{equation}
In the degenerate case write $I$ for the set of $D_{N}\in[0,1]$ with
$a_{n}+b_{n}D_{N}\ge0$ for every $n<N$.

\begin{proposition}[Feasibility]
\label{prop:feasibility}
Let Assumption~\ref{ass:nonzero} hold and fix a reference index $N$. Exactly one
of the following holds.
\begin{enumerate}
\item[(i)] If $\sigma\neq0$, equivalently $1+\sum_{n<N}b_{n}\neq0$, there is a
unique vector satisfying \eqref{eq:indiff} and $\sum_{n}D_{n}=1$, namely
\eqref{eq:Dv} with $v=v^{*}:=(\rho-1)/\sigma$, equivalently \eqref{eq:affine}
with $D_{N}=D_{N}^{*}$. It is a probability vector if and only if
\begin{equation}
D_{N}^{*}\ge0\quad\text{and}\quad a_{n}+b_{n}D_{N}^{*}\ge0\ \ (1\le n\le N-1).
\label{eq:feas}
\end{equation}
\item[(ii)] If $\sigma=0$ and $\rho\neq1$, no normalised vector on the locus
exists.
\item[(iii)] If $\sigma=0$ and $\rho=1$, normalisation holds identically along
the locus and the set of valid protection vectors is parametrised by
$D_{N}\in I$, a closed subinterval of $[0,1]$, possibly empty or a single point.
\end{enumerate}
\end{proposition}

\begin{proof}
By \eqref{eq:identities}, $1+\sum_{n<N}b_{n}=\Omega_{N}^{\R}\sigma$ with
$\Omega_{N}^{\R}\ne0$, so the two conditions in (i) agree and normalisation is
the scalar equation \eqref{eq:scalar}. If $\sigma\ne0$ this has the unique root
$v^{*}$, giving a unique $D$ by \eqref{eq:Dv}; comparison with \eqref{eq:norm}
gives $D_{N}^{*}$. If $\sigma=0$, \eqref{eq:scalar} reads $0=\rho-1$, which is
unsolvable for $\rho\ne1$ and holds for every $v$ when $\rho=1$.

For feasibility in (i): normalisation holds by construction, so if each
$D_{n}\ge0$ then each $D_{n}\le1$ automatically; the nonnegativity conditions
\eqref{eq:feas} are therefore necessary and sufficient and the upper bounds need
not be imposed. The same applies pointwise in (iii), so $I$ is exactly the
feasible set. Each constraint $a_{n}+b_{n}D_{N}\ge0$ defines a closed half-line
or all of $\mathbb{R}$; $I$ is the intersection of finitely many such sets with
$[0,1]$, hence a closed subinterval of $[0,1]$.
\end{proof}

\begin{proposition}[Sign dichotomy]
\label{prop:signdichotomy}
If the payoff swings $\Omega_{n}^{\R}$, $n=1,\ldots,N$, all have the same sign
then $\sigma\neq0$, so case (iii) of Proposition~\ref{prop:feasibility} cannot
occur and the locus meets the normalisation hyperplane in at most one point.
Consequently a nondegenerate feasible interval of full-support protection
vectors can exist only if $\Omega_{j}^{\R}>0>\Omega_{k}^{\R}$ for some $j,k$.
\end{proposition}

\begin{proof}
If $\Omega_{n}^{\R}>0$ for all $n$ then every summand in \eqref{eq:sigmarho} is
positive, so $\sigma>0$; if $\Omega_{n}^{\R}<0$ for all $n$ then $\sigma<0$.
Either way Proposition~\ref{prop:feasibility}\,(i) applies, and the last
statement is the contrapositive.
\end{proof}

\begin{corollary}[Sign-regular case]
\label{cor:signregular}
Assume $\Omega_{n}^{\R}>0$ for all $n$. Then $\sigma>0$, the unique candidate
attacker payoff is $v^{*}=(\rho-1)/\sigma$, and the full-support construction is
feasible if and only if
\begin{equation}
v^{*}\le\min_{1\le n\le N}\RR(\Tk{n}),
\label{eq:cleanfeas}
\end{equation}
in which case the protection vector is
$D_{n}=(\RR(\Tk{n})-v^{*})/\Omega_{n}^{\R}$.
\end{corollary}

\begin{proof}
By \eqref{eq:Dv}, $D_{n}\ge0$ is equivalent to $\RR(\Tk{n})\ge v^{*}$ when
$\Omega_{n}^{\R}>0$; requiring this for all $n$ gives \eqref{eq:cleanfeas}. The
upper bounds are automatic by Proposition~\ref{prop:feasibility}.
\end{proof}

Condition \eqref{eq:cleanfeas} states that the full-support configuration is
available precisely when the attacker's normalisation-implied payoff does not
exceed the smallest unprotected reward in the target set. If it does, the
defender cannot make every target simultaneously unattractive with a unit
coverage budget and the equilibrium has boundary support.

\section{Strong Stackelberg tie-breaking}
\label{sec:envelope}

On the full-support locus every target yields the attacker the same payoff, so
by \eqref{eq:BR}
\begin{equation}
BR_{\R}(D)=\simp_{N}:
\label{eq:BRfull}
\end{equation}
the follower's best-response set is the entire attack simplex. Writing $g_{n}$
as a function of the free parameter,
\begin{equation}
g_{n}(D_{N})=D_{n}(D_{N})\Omega_{n}^{\B}-\CB(\Tk{n}),\qquad n=1,\ldots,N,
\label{eq:gn_param}
\end{equation}
with $D_{n}(D_{N})=a_{n}+b_{n}D_{N}$ for $n<N$, the defender's payoff under
\eqref{eq:qW} at a feasible $D_{N}$ is
\begin{equation}
G(D_{N})=\max_{A\in\simp_{N}}\sum_{n=1}^{N}A_{n}g_{n}(D_{N})
=\max_{1\le n\le N}g_{n}(D_{N}),
\label{eq:envelope}
\end{equation}
the selected attack being concentrated on any maximising target.

\begin{proposition}[Endpoint selection]
\label{prop:endpoint}
Suppose Proposition~\ref{prop:feasibility}\,(iii) applies with feasible interval
$I=[L,U]\subseteq[0,1]$, and let $g_{n}$, $G$ be as in
\eqref{eq:gn_param}--\eqref{eq:envelope}. Then $G$ is convex and piecewise
affine on $I$ and
\begin{equation}
\max_{D_{N}\in I}G(D_{N})=\max\{G(L),\,G(U)\}.
\label{eq:endpointrule}
\end{equation}
If $\widehat{D}_{N}$ denotes an endpoint attaining the maximum in
\eqref{eq:endpointrule}, the associated attack vector may be taken to be a point
mass on any $n\in\arg\max_{k}g_{k}(\widehat{D}_{N})$. (The symbol $D_{N}^{*}$ of
\eqref{eq:DNstar} is reserved for the unique candidate of
Proposition~\ref{prop:feasibility}\,(i) and does not apply here.)
\end{proposition}

\begin{proof}
Each $g_{n}$ in \eqref{eq:gn_param} is affine in $D_{N}$, so $G$ is a pointwise
maximum of finitely many affine functions, hence convex and piecewise affine. A
convex function on a compact interval attains its maximum at an extreme point.
\end{proof}

By Proposition~\ref{prop:signdichotomy} the hypothesis of
Proposition~\ref{prop:endpoint} can be met only under sign-irregular parameters;
otherwise $D_{N}$ is determined uniquely and no endpoint comparison arises.
Under a weak Stackelberg convention the maximum in \eqref{eq:envelope} is
replaced by a minimum, $\min_{n}g_{n}$ is concave, and the maximum over $I$ is
generally interior.

Proposition~\ref{prop:endpoint} optimises over the full-support locus only, so
global optimality requires separate verification. Throughout we use for that
purpose the target-conditional programmes of \citet{ConitzerSandholm2006}: for
each target $t$,
\begin{equation}
\max_{D\in\mathcal{D}} D_{t}\Omega_{t}^{\B}-\CB(\Tk{t})
\quad\text{s.t.}\quad
\RR(\Tk{t})-D_{t}\Omega_{t}^{\R}\ \ge\ \RR(\Tk{k})-D_{k}\Omega_{k}^{\R}
\ \ \forall k,
\label{eq:LPt}
\end{equation}
where $\mathcal{D}$ is the defender's feasible set. The feasible set of
\eqref{eq:LPt} is the region of $\mathcal{D}$ on which $t$ is a best response,
its value is the best defender payoff obtainable while $t$ is attacked, and the
SSE value is the maximum of the $N$ programme values.

\section{Extension: a general coverage budget}
\label{sec:budget}

The unit-allocation simplex fixes total coverage at one.  We now pass to
$M$ identical resources and the marginal-coverage interpretation introduced in
Section~\ref{sec:model}.  Fix $1\le M<N$ and allow resources to be idle, so
the defender's feasible set is
\begin{equation}
\mathcal{D}_{N,M}=\Big\{D\in\mathbb{R}^{N}:\ 0\le D_{n}\le1\ \ \forall n,\quad
\sum_{n=1}^{N}D_{n}\le M\Big\},
\label{eq:DNM}
\end{equation}
the convex hull of incidence vectors of target subsets of cardinality at most
$M$.  Hence it is exactly the marginal-coverage set implementable by $M$
identical resources when no scheduling constraints are present and idling is
allowed \citep{Kiekintveld2009,Korzhyk2011}. The payoff functions, the attacker's problem
and the indifference algebra of Section~\ref{sec:indifference} are unchanged;
only the feasibility step differs. Because the free parameter is now the scalar
$v$ rather than a coordinate of $D$, the reference-free description of
Lemma~\ref{lem:symmetric} carries the whole analysis.

\subsection{Feasibility}
\label{sec:budget_feas}

For each target set
\begin{equation}
J_{n}=\Big[\min\{\RR(\Tk{n}),-\CR(\Tk{n})\},\
\max\{\RR(\Tk{n}),-\CR(\Tk{n})\}\Big].
\label{eq:Jn}
\end{equation}

\begin{lemma}
\label{lem:box}
Let Assumption~\ref{ass:nonzero} hold and $D_{n}(v)=(\RR(\Tk{n})-v)/
\Omega_{n}^{\R}$. Then $0\le D_{n}(v)\le1$ if and only if $v\in J_{n}$, with
$D_{n}(v)=0$ at the endpoint $\RR(\Tk{n})$ and $D_{n}(v)=1$ at the endpoint
$-\CR(\Tk{n})$.
\end{lemma}

\begin{proof}
If $\Omega_{n}^{\R}>0$ then $D_{n}(v)\ge0$ iff $v\le\RR(\Tk{n})$, and
$D_{n}(v)\le1$ iff $\RR(\Tk{n})-v\le\Omega_{n}^{\R}$, that is
$v\ge-\CR(\Tk{n})$; since $\Omega_{n}^{\R}>0$ gives
$\RR(\Tk{n})>-\CR(\Tk{n})$, the two conditions together read $v\in
[-\CR(\Tk{n}),\RR(\Tk{n})]=J_{n}$. If $\Omega_{n}^{\R}<0$ both inequalities
reverse and $\RR(\Tk{n})<-\CR(\Tk{n})$, giving $v\in J_{n}$ again. Substituting
the endpoints into $D_{n}(v)$ gives the stated values.
\end{proof}

\begin{theorem}[Feasibility under a coverage budget]
\label{thm:budget}
Let $B_{M}=\{v\in\mathbb{R}:\ \rho-\sigma v\le M\}$ and
\begin{equation}
V_{M}=\Big(\bigcap_{n=1}^{N}J_{n}\Big)\cap B_{M}.
\label{eq:VM}
\end{equation}
Then $D\in\mathcal{D}_{N,M}$ lies on the full-support indifference locus if and
only if $D=D(v)$ for some $v\in V_{M}$. The set $B_{M}$ is a closed half-line
when $\sigma\neq0$, all of $\mathbb{R}$ when $\sigma=0$ and $\rho\le M$, and
empty when $\sigma=0$ and $\rho>M$; consequently $V_{M}$ is a closed interval,
possibly empty or a single point.
\end{theorem}

\begin{proof}
By Lemma~\ref{lem:symmetric} the locus is $\{D(v):v\in\mathbb{R}\}$ and
$\sum_{n}D_{n}(v)=\rho-\sigma v$, so the budget constraint is $v\in B_{M}$; by
Lemma~\ref{lem:box} the box constraints are $v\in\bigcap_{n}J_{n}$. The
description of $B_{M}$ is immediate from the sign of $\sigma$. Each $J_{n}$ is a
closed bounded interval, so $\bigcap_{n}J_{n}$ is one, and its intersection with
a closed half-line or with $\mathbb{R}$ is a closed interval.
\end{proof}

\begin{corollary}[Sign-regular case]
\label{cor:budget_regular}
If $\Omega_{n}^{\R}>0$ for all $n$ then $\sigma>0$ and $V_{M}=[v_{-},v_{+}]$
with
\begin{equation}
v_{-}=\max\Big\{\frac{\rho-M}{\sigma},\ -\min_{n}\CR(\Tk{n})\Big\},
\qquad
v_{+}=\min_{n}\RR(\Tk{n}),
\label{eq:vpm}
\end{equation}
nonempty precisely when $v_{-}\le v_{+}$. At $v=(\rho-M)/\sigma$ the budget is
exactly exhausted, $\sum_{n}D_{n}=M$; at $v=-\CR(\Tk{j})$ the target $\Tk{j}$ is
fully covered, $D_{j}=1$; and at $v=\RR(\Tk{j})$ it is left uncovered,
$D_{j}=0$.
\end{corollary}

\begin{proof}
Positivity of $\sigma$ follows from Proposition~\ref{prop:signdichotomy}.
When $\Omega_{n}^{\R}>0$, Lemma~\ref{lem:box} gives
$J_{n}=[-\CR(\Tk{n}),\RR(\Tk{n})]$, and hence
\[
\bigcap_nJ_n=[-\min_n\CR(\Tk{n}),\min_n\RR(\Tk{n})].
\]
Moreover, $\sigma>0$ makes $B_M$ the half-line
$v\ge(\rho-M)/\sigma$. Substituting
$v=(\rho-M)/\sigma$ into $\sum_{n}D_{n}(v)=\rho-\sigma v$ gives $M$; the other
two statements are the endpoint values of Lemma~\ref{lem:box}.
\end{proof}

Imposing $M=1$ with equality in place of the inequality recovers
Proposition~\ref{prop:feasibility}: the budget constraint becomes
$\sigma v=\rho-1$, which has a unique root when $\sigma\ne0$ and is either
vacuous or unsolvable when $\sigma=0$. The unit-budget analysis is therefore the
corner of the present one in which the budget binds exactly, and the trichotomy
of Proposition~\ref{prop:feasibility} is an artefact of replacing an inequality
by an equality: under \eqref{eq:DNM} a nondegenerate feasible interval is
generic, and by Corollary~\ref{cor:budget_regular} it occurs under sign-regular
parameters, where Proposition~\ref{prop:signdichotomy} had excluded it.

\subsection{Selection}
\label{sec:budget_selection}

On the locus the defender's pure-target payoffs are, as functions of $v$,
\begin{equation}
\gamma_{n}(v)=\frac{\RR(\Tk{n})-v}{\Omega_{n}^{\R}}\Omega_{n}^{\B}-\CB(\Tk{n}),
\qquad
\Gamma(v)=\max_{1\le n\le N}\gamma_{n}(v).
\label{eq:gamma}
\end{equation}

\begin{proposition}[Endpoint selection under a budget]
\label{prop:budget_endpoint}
$\Gamma$ is convex and piecewise affine, and if $V_{M}=[v_{-},v_{+}]$ is
nonempty then
\begin{equation}
\max_{v\in V_{M}}\Gamma(v)=\max\{\Gamma(v_{-}),\,\Gamma(v_{+})\}.
\label{eq:budget_endpoint}
\end{equation}
\end{proposition}

\begin{proof}
Each $\gamma_{n}$ is affine in $v$ with slope
$-\Omega_{n}^{\B}/\Omega_{n}^{\R}$, so $\Gamma$ is a pointwise maximum of
finitely many affine functions and a convex function on a compact interval
attains its maximum at an extreme point.
\end{proof}

\begin{proposition}[Monotone case]
\label{prop:budget_monotone}
If $\Omega_{n}^{\R}>0$ and $\Omega_{n}^{\B}>0$ for every $n$, then every
$\gamma_{n}$ is strictly decreasing, hence so is $\Gamma$, and the maximum over
$V_{M}$ is attained at $v_{-}$. By Corollary~\ref{cor:budget_regular} the
defender therefore either exhausts the budget, $\sum_{n}D_{n}=M$, or covers some
target fully, $D_{j}=1$.
\end{proposition}

\begin{proof}
The slope of $\gamma_{n}$ is $-\Omega_{n}^{\B}/\Omega_{n}^{\R}<0$ under the
stated signs, and a maximum of strictly decreasing functions is strictly
decreasing.
\end{proof}

Under sign regularity on both players' swings,
Proposition~\ref{prop:budget_monotone} places a full-support optimum at the
lowest feasible attacker value, where by Corollary~\ref{cor:budget_regular}
either the coverage budget is exhausted or some target is fully covered. In that
regime the defender never withholds coverage without saturating a target.

Coverage left unused with no target saturated therefore requires one of the two
sign hypotheses to fail, and it can fail on either side. A negative
defender-side swing $\Omega_{n}^{\B}<0$ destroys the monotonicity of $\Gamma$
and admits an interior or upper-endpoint optimum. A negative attacker-side swing
is equally sufficient, and it is the mechanism that operates in
Section~\ref{sec:dynamic}: if some $\Omega_{n}^{\R}<0$ then $\sigma$ may be
negative, in which case the budget half-line $B_{M}$ of
Theorem~\ref{thm:budget} bounds $v$ from \emph{above} rather than below, so the
budget is slack at the selected lower endpoint while that endpoint is fixed by a
box constraint. Section~\ref{sec:dynamic} generates precisely this
configuration, endogenously and from strictly positive primitive swings on both
sides.

Saturation of an individual cap is a distinct phenomenon, and is compatible with
full sign regularity: if $-\min_{n}\CR(\Tk{n})>(\rho-M)/\sigma$ then $v_{-}$ is
set by a cap, some $D_{j}=1$, and the budget is slack with no negative swing
anywhere. Proposition~\ref{prop:budget_monotone} also concerns the full-support
locus alone; boundary-support optima are governed by the conditional programmes
\eqref{eq:LPt}.

\section{A dynamic extension: attrition and continuation-adjusted
indifference}
\label{sec:dynamic}

Successful attacks now remove assets, so current coverage changes the future
target set. We show that continuation-adjusted payoffs preserve the static
node structure and can endogenously create the sign reversals excluded by the
primitive sign-regular model.

\subsection{The attrition game}
\label{sec:attrition}

Let $H\in\mathbb{N}$ be the horizon and $\delta\in[0,1]$ a discount factor. The
state is the set $S\subseteq\{1,\ldots,N\}$ of surviving assets. In each period,
$\B$ commits to a coverage vector $D\in\mathcal{D}_{|S|,M}$ on $S$, $\R$ observes
it and strikes. A strike on a covered target leaves the target intact and yields
the period payoffs $\RB(\Tk{n})$ to $\B$ and $-\CR(\Tk{n})$ to $\R$; a strike on
an uncovered target destroys it, removing $n$ from the state, and yields
$-\CB(\Tk{n})$ and $\RR(\Tk{n})$. Play stops after $H$ periods or when
$S=\emptyset$, at which point both continuation values are zero.

The leader recommits after each observed state, so the solution is a stagewise
feedback-recommitment equilibrium evaluated by backward induction, with
\eqref{eq:lexSSE} imposed at every node. It is not an initial commitment to an
entire contingent policy.

Write $V_{\B}^{t}(S)$ and $V_{\R}^{t}(S)$ for the values with $t$ periods
remaining, and $S\!-\!n$ for $S\setminus\{n\}$. When $M\ge|S|$ the budget in
\eqref{eq:DNM} is inactive and the defender can cover $S$ entirely, so we read
the budget as $\min\{M,|S|\}$ throughout. Conditioning on the strike
target and on whether it is covered,
\begin{align}
u_{n}(D)&=(1-D_{n})\big[\RR(\Tk{n})+\delta V_{\R}^{t-1}(S\!-\!n)\big]
+D_{n}\big[-\CR(\Tk{n})+\delta V_{\R}^{t-1}(S)\big],
\label{eq:dyn_u}\\
g_{n}(D)&=D_{n}\big[\RB(\Tk{n})+\delta V_{\B}^{t-1}(S)\big]
-(1-D_{n})\big[\CB(\Tk{n})-\delta V_{\B}^{t-1}(S\!-\!n)\big].
\label{eq:dyn_g}
\end{align}

\begin{lemma}[Reduction to the static form]
\label{lem:reduction}
Define the continuation-adjusted parameters
\begin{equation}
\begin{aligned}
\widetilde{\RR}(\Tk{n})&=\RR(\Tk{n})+\delta V_{\R}^{t-1}(S\!-\!n), &\qquad
\widetilde{\CR}(\Tk{n})&=\CR(\Tk{n})-\delta V_{\R}^{t-1}(S),\\
\widetilde{\RB}(\Tk{n})&=\RB(\Tk{n})+\delta V_{\B}^{t-1}(S), &\qquad
\widetilde{\CB}(\Tk{n})&=\CB(\Tk{n})-\delta V_{\B}^{t-1}(S\!-\!n).
\end{aligned}
\label{eq:tilde}
\end{equation}
Then \eqref{eq:dyn_u} and \eqref{eq:dyn_g} coincide with the static per-target
payoffs \eqref{eq:un_gn} built from $\widetilde{\RR},\widetilde{\CR},
\widetilde{\RB},\widetilde{\CB}$. In particular the adjusted swings are
\begin{equation}
\widetilde{\Omega}_{n}^{\R}=\Omega_{n}^{\R}
+\delta\big[V_{\R}^{t-1}(S\!-\!n)-V_{\R}^{t-1}(S)\big],
\qquad
\widetilde{\Omega}_{n}^{\B}=\Omega_{n}^{\B}
+\delta\big[V_{\B}^{t-1}(S)-V_{\B}^{t-1}(S\!-\!n)\big].
\label{eq:tilde_omega}
\end{equation}
\end{lemma}

\begin{proof}
Substituting \eqref{eq:tilde} into \eqref{eq:dyn_u} gives
$u_{n}(D)=(1-D_{n})\widetilde{\RR}(\Tk{n})-D_{n}\widetilde{\CR}(\Tk{n})$, and
into \eqref{eq:dyn_g} gives
$g_{n}(D)=D_{n}\widetilde{\RB}(\Tk{n})-(1-D_{n})\widetilde{\CB}(\Tk{n})$, which
are \eqref{eq:PiB}--\eqref{eq:PiR} target by target. Adding the two adjusted
parameters in each pair gives \eqref{eq:tilde_omega}.
\end{proof}

\begin{corollary}[Nodewise computation]
\label{cor:statewise}
At every state--period node, Lemma~\ref{lem:reduction} produces a static game
of the form analysed in Sections~\ref{sec:indifference}--\ref{sec:budget}.
The full-support formulae and endpoint rule therefore apply after replacing the
primitive parameters by their tilde counterparts.  If the
continuation-adjusted full-support locus is feasible and nondegenerate, the best
point on that locus is obtained from the scalar interval and its endpoints.  A
global nodewise SSE is obtained from that closed-form candidate only after it
is compared with any relevant boundary-support solutions; when the locus is
infeasible, target-conditional linear programmes are required.
\end{corollary}

\begin{proof}
Lemma~\ref{lem:reduction} identifies the node's payoff functions with the static
ones.  The full-support statements then follow from
Lemma~\ref{lem:symmetric}, Theorem~\ref{thm:budget} and
Proposition~\ref{prop:budget_endpoint}.  Global SSE computation must also allow
the follower's support to be a proper subset, which is exactly what the
conditional programmes \eqref{eq:LPt} enumerate.
\end{proof}

The refinement makes the recursion single-valued: maximise the defender value,
then minimise the associated attacker value, and finally choose a resource-
minimal representative. A direct implementation has at most $H(2^N-1)$
nonempty nodes; full-support nodes use scalar calculations plus a global support
check, while a boundary node may require up to $|S|$ conditional programmes.

\subsection{Endogenous sign reversal}
\label{sec:reversal}

Equation \eqref{eq:tilde_omega} has an immediate consequence.

\begin{proposition}[Endogenous sign reversal]
\label{prop:reversal}
Suppose $\Omega_{n}^{\R}>0$, so that coverage of $\Tk{n}$ is immediately costly
to the attacker. Then $\widetilde{\Omega}_{n}^{\R}<0$ if and only if
\begin{equation}
\delta\big[V_{\R}^{t-1}(S)-V_{\R}^{t-1}(S\!-\!n)\big]>\Omega_{n}^{\R},
\label{eq:reversal}
\end{equation}
that is, if and only if the attacker's continuation loss from destroying
$\Tk{n}$ exceeds its immediate gain from doing so. In that case coverage of
$\Tk{n}$ strictly \emph{raises} the attacker's payoff at $\Tk{n}$, because a
covered strike leaves the asset available for later exploitation.
\end{proposition}

\begin{proof}
Immediate from \eqref{eq:tilde_omega}.
\end{proof}

\begin{remark}[No monotonicity assumption]
\label{rem:direction}
Equation~\eqref{eq:tilde_omega} does not by itself determine the sign of the
continuation adjustment: it depends on how each player's equilibrium value
changes when a target is removed.  No monotonicity under set inclusion is
assumed or required in the results.  In particular, the attacker has no option
to abstain, so even $V_{\R}^{t}(\{n\})$ may be below
$V_{\R}^{t}(\emptyset)=0$, while the defender can be harmed by additional assets
because they dilute a fixed coverage budget.  Proposition~\ref{prop:reversal}
is therefore stated directly in terms of the relevant continuation-value
difference rather than through a monotonicity premise.
\end{remark}

Thus sign irregularity need not be imposed on primitive payoffs: it arises when
an asset is cheap to destroy now but valuable to preserve for later attacks. On
the defender side, $\widetilde{\Omega}_{n}^{\B}<0$ similarly removes the
monotonicity behind Proposition~\ref{prop:budget_monotone} and can make unused
budget optimal, subject to the attacker's global best-response constraints.

\subsection{A three-asset, two-period instance}
\label{sec:dyn_example}

\begin{example}
\label{ex:four}
Let $N=3$, $M=1$, $H=2$, $\delta=1$, with the parameters of
Table~\ref{tab:ex4}. All are nonnegative and all static swings are strictly
positive, so the primitives are fully sign-regular in the sense of
Proposition~\ref{prop:budget_monotone}.
\end{example}

\begin{table}[t]
\centering
\caption{Primitive payoff parameters for Example~\ref{ex:four}.}
\label{tab:ex4}
\begin{tabular}{crrrrrr}
\toprule
Asset & $\RB$ & $\CB$ & $\RR$ & $\CR$ & $\Omega^{\B}$ & $\Omega^{\R}$\\
\midrule
$\Tk{1}$ & $5$ & $6$ & $1$ & $0$ & $11$ & $1$\\
$\Tk{2}$ & $5$ & $6$ & $4$ & $7$ & $11$ & $11$\\
$\Tk{3}$ & $6$ & $1$ & $4$ & $7$ & $7$  & $11$\\
\bottomrule
\end{tabular}
\end{table}

In the terminal period the continuation values vanish and each node is the
static game of Section~\ref{sec:budget}. Solving the seven nonempty states,
\begin{equation}
\begin{aligned}
V_{\R}^{1}(\{1\})&=0, & V_{\R}^{1}(\{2\})&=V_{\R}^{1}(\{3\})=-7, &
V_{\R}^{1}(\{1,2\})&=V_{\R}^{1}(\{1,3\})=\tfrac13,\\
V_{\R}^{1}(\{2,3\})&=-\tfrac32, & V_{\R}^{1}(\{1,2,3\})&=\tfrac8{13}, &&
\end{aligned}
\label{eq:ex4_VR}
\end{equation}
with $V_{\B}^{1}(\{1,2,3\})=\tfrac{15}{13}$ attained at
$D=\langle\tfrac5{13},\tfrac4{13},\tfrac4{13}\rangle$ and a tie-break on
$\Tk{3}$.

Now pass to the first period at the full state $S=\{1,2,3\}$. By
\eqref{eq:tilde_omega},
\begin{equation}
\widetilde{\Omega}_{1}^{\R}=1+\Big(-\tfrac32-\tfrac8{13}\Big)=-\tfrac{29}{26}<0,
\qquad
\widetilde{\Omega}_{2}^{\R}=\widetilde{\Omega}_{3}^{\R}
=11-\tfrac{11}{39}=\tfrac{418}{39}>0.
\label{eq:ex4_omt}
\end{equation}
The swing at $\Tk{1}$ has changed sign: destroying $\Tk{1}$ costs the attacker
$\tfrac{55}{26}$ in continuation value against an immediate swing of only
$\Omega_{1}^{\R}=1$, so condition \eqref{eq:reversal} holds. The first-period
solution is full-support, with common attacker payoff
$\widetilde{v}=-\tfrac12$ and
\begin{equation}
D^{*}=\Big\langle 0,\ \tfrac{377}{836},\ \tfrac{377}{836}\Big\rangle,
\qquad
V_{\B}^{2}(\{1,2,3\})=\tfrac{75}{22},
\qquad
V_{\R}^{2}(\{1,2,3\})=-\tfrac12,
\label{eq:ex4_sol}
\end{equation}
with the tie-break on $\Tk{3}$.

\begin{remark}[The forward-looking defender abandons an asset]
\label{rem:abandon}
Although a myopic defender assigns the largest share to $\Tk{1}$,
\eqref{eq:ex4_sol} sets $D_1=0$ and leaves budget unused. Because
$\widetilde{\Omega}_{1}^{\R}<0$, covering $\Tk{1}$ raises its immediate attack
value by preserving it for future strikes. Withdrawing coverage is therefore
both less protective and more deterrent: continuation values separate the two
roles.
\end{remark}

\begin{remark}[Dependence on the absence of an attacker outside option]
\label{rem:outside}
The attacker in Section~\ref{sec:attrition} strikes some surviving target in
every period, and the mechanism depends on that assumption. Suppose instead the
attacker may decline to strike, receiving $0$ in the current period and leaving
the state unchanged.
Then $V_{\R}^{t}(S)\ge0$ for every $t$ and $S$ by backward induction, so the
continuation gap $V_{\R}^{t-1}(S)-V_{\R}^{t-1}(S\!-\!n)$ in
\eqref{eq:tilde_omega} is compressed from below, and condition
\eqref{eq:reversal} becomes correspondingly harder to satisfy.

In Example~\ref{ex:four} the effect disappears entirely. With an outside option
the terminal values become $V_{\R}^{1}(\{2\})=V_{\R}^{1}(\{3\})=
V_{\R}^{1}(\{2,3\})=0$, in place of $-7$, $-7$ and $-3/2$, because the attacker
declines a strike rather than accept a negative payoff. The adjusted swing at
$\Tk{1}$ is then $\widetilde{\Omega}_{1}^{\R}=395/2912>0$ rather than
$-29/26<0$, the first-period selection is
$\langle\tfrac{79}{224},\tfrac{145}{448},\tfrac{145}{448}\rangle$ with the
budget binding, and neither abandonment nor unused budget occurs. A randomised
search over several thousand two-period, three-asset instances with an outside
option, concentrated on the parameter region most favourable to reversal, did
not produce a single negative adjusted attacker swing.

Forced engagement is natural in the combat setting that motivates the model,
where the attacking force is committed, and it is implicit in one-shot security
games, where the attacker's participation is not at issue. It is a stronger
assumption in domains such as poaching or illegal fishing, where declining to
act is both available and cheap. Proposition~\ref{prop:reversal} is valid as
derived, but the instances exhibited here rely on the attacker having no outside
option; conditions under which reversal survives such an option remain open.
\end{remark}

Example~\ref{ex:four} also illustrates the practical content of
Corollary~\ref{cor:statewise}: the two-period, three-asset game requires seven terminal-state evaluations and
one first-period full-state evaluation; the full-support formula applies at the
reported nodes, while the conditional LP formulation supplies an independent
global check.

\subsection{Comparative statics and the singular discount threshold}
\label{sec:sweep}

The terminal values do not depend on $\delta$, so the first-period adjusted
attacker swings are affine:
\begin{equation}
\widetilde{\Omega}_{1}^{\R}(\delta)=1-\tfrac{55}{26}\delta,
\qquad
\widetilde{\Omega}_{2}^{\R}(\delta)=\widetilde{\Omega}_{3}^{\R}(\delta)
=11-\tfrac{11}{39}\delta .
\label{eq:omt_delta}
\end{equation}
Only the first can vanish on $[0,1]$, at
\begin{equation}
\delta^{*}=\tfrac{26}{55}\approx0.4727.
\label{eq:dstar}
\end{equation}
Away from $\delta^*$, the nonzero-swing full-support formula is available.  At
$\delta^*$ the formula must not be used, but the node itself remains a regular
linear Stackelberg problem and can be solved exactly.

\begin{proposition}[Exact equilibrium at the singular node]
\label{prop:threshold}
At $\delta=\delta^*$, the first-period full-state node has adjusted attacker
payoffs
\begin{equation}
u_1(D)=\tfrac{16}{55},\qquad
u_2(D)=\tfrac{686}{165}-\tfrac{163}{15}D_2,
\qquad
u_3(D)=\tfrac{686}{165}-\tfrac{163}{15}D_3.
\label{eq:threshold_u}
\end{equation}
Its global strong Stackelberg defender value is $23/11$ and its follower value is
$16/55$.  The defender-optimal set is exactly the triangle
\begin{equation}
\Big\{D:\ D_3=\tfrac{58}{163},\quad D_2\ge\tfrac{58}{163},\quad D_1\ge0,
\quad D_1+D_2\le\tfrac{105}{163}\Big\},
\label{eq:threshold_set}
\end{equation}
whose vertices are $\langle0,\tfrac{58}{163},\tfrac{58}{163}\rangle$,
$\langle\tfrac{47}{163},\tfrac{58}{163},\tfrac{58}{163}\rangle$ and
$\langle0,\tfrac{105}{163},\tfrac{58}{163}\rangle$.  It is two-dimensional: the
defender payoff at the struck target depends only on $D_3$, so $D_1$ and $D_2$
are free subject to keeping $\Tk{3}$ a best response and respecting the budget.
Under the resource-minimal refinement \eqref{eq:lexSSE}, the selected coverage is
\begin{equation}
D^*(\delta^*)=\big\langle0,\tfrac{58}{163},\tfrac{58}{163}\big\rangle,
\qquad \sum_nD_n^*(\delta^*)=\tfrac{116}{163}.
\label{eq:threshold_D}
\end{equation}
\end{proposition}

\begin{proof}
At $\delta^*$, $\widetilde{\Omega}_1^{\R}=0$, while equalising targets 2 and 3
with $u_1=16/55$ gives $D_2=D_3=58/163$. The three conditional programmes then
have values $-3281/1793$, $-2664/1793$, and $23/11$ for targets 1, 2, and 3,
respectively. For target 3, optimality requires $D_3=58/163$, and its
best-response and budget constraints reduce to
$D_2\ge58/163$, $D_1\ge0$, and $D_1+D_2\le105/163$, yielding exactly
\eqref{eq:threshold_set}. The attacker value is constant at $16/55$ on this set;
minimising total coverage selects \eqref{eq:threshold_D}.
\end{proof}

The optimal face appears because $\widetilde{\Omega}_1^{\R}=0$ makes coverage of
$\Tk{1}$ irrelevant to the attacker's payoff. Equation~\eqref{eq:lexSSE}
selects one representative from that face.

\begin{theorem}[Global discount-factor equilibrium]
\label{thm:discount_global}
For the first-period full-state node of Example~\ref{ex:four}, target $\Tk{3}$
is selected by strong tie-breaking for every $\delta\in[0,1]$.  Writing
$\delta^*=26/55$, the lexicographically selected coverage and the two values are
as follows.

If $0\le\delta<\delta^*$,
\begin{equation}
D^*(\delta)=\left\langle
\frac{195-154\delta}{507-176\delta},
\frac{156-11\delta}{507-176\delta},
\frac{156-11\delta}{507-176\delta}
\right\rangle,
\label{eq:policy_left}
\end{equation}
\begin{equation}
V_{\B}^{2}=\frac{7605+9711\delta-3025\delta^2}
 {13(507-176\delta)},\qquad
V_{\R}^{2}=\frac{4056-4810\delta-803\delta^2}
 {13(507-176\delta)}.
\label{eq:values_left}
\end{equation}
The budget binds.  At $\delta=\delta^*$, Proposition~\ref{prop:threshold}
applies.  If $\delta^*<\delta\le1$,
\begin{equation}
D^*(\delta)=\left\langle 0,
\frac{13(11\delta+18)}{22(39-\delta)},
\frac{13(11\delta+18)}{22(39-\delta)}
\right\rangle,
\label{eq:policy_right}
\end{equation}
\begin{equation}
V_{\B}^{2}=\frac{5(11\delta+4)}{22},\qquad
V_{\R}^{2}=1-\frac32\delta .
\label{eq:values_right}
\end{equation}
The budget is slack.  Both value functions are continuous at $\delta^*$, while
the selected policy is discontinuous because the resource-minimal refinement
chooses $D_1=0$ at the singular node.
\end{theorem}

\begin{proof}
Let
\begin{equation*}
r=1-\frac{55}{26}\delta,\qquad
b=11-\frac{11}{39}\delta,\qquad
\Delta=\widetilde{\RR}_2-\widetilde{\RR}_1
      =\frac{11\delta+18}{6}.
\end{equation*}
The attacker payoffs can be written
$u_1=\widetilde{\RR}_1-rD_1$ and
$u_2=\widetilde{\RR}_2-bD_2$,
$u_3=\widetilde{\RR}_2-bD_3$.
For target 3 to be a best response, its constraints are
$D_3\le D_2$ and $bD_3-rD_1\le\Delta$.  The target-3 defender objective is
strictly increasing in $D_3$ throughout $[0,1]$.

For target 3, any excess of $D_2$ over $D_3$ uses budget without relaxing
$bD_3-rD_1\le\Delta$, so an optimum can be chosen with $D_2=D_3$. When $r>0$,
increasing $D_1$ relaxes that constraint. Both it and the budget therefore bind,
which gives
\begin{equation*}
D_2=D_3=\frac{\Delta+r}{b+2r}
       =\frac{156-11\delta}{507-176\delta},\qquad
D_1=\frac{b-2\Delta}{b+2r}
       =\frac{195-154\delta}{507-176\delta}.
\end{equation*}
These components lie in $[0,1]$ for $0\le\delta<\delta^*$. When $r<0$,
positive $D_1$ both uses budget and tightens the attacker constraint, so
$D_1=0$. Maximising $D_3$ then gives $D_2=D_3=\Delta/b$, which is
\eqref{eq:policy_right}; moreover $0<\Delta/b<1/2$ for
$\delta\in(\delta^*,1]$, so all component bounds hold and the budget is slack.
The target-2 conditional programme is obtained by exchanging targets 2 and 3,
so it has the same optimal coverage vector, although its defender objective has
a different value.

For target 1 to be a best response, the constraints are
$bD_2-rD_1\ge\Delta$ and $bD_3-rD_1\ge\Delta$. For fixed $D_1$, the least
coverage consistent with those constraints has
$D_2=D_3=(\Delta+rD_1)/b$. Since the target-1 defender objective is increasing
in $D_1$, the budget binds, and solving
$D_1+2(\Delta+rD_1)/b=1$ gives
\begin{equation*}
D_1=\frac{b-2\Delta}{b+2r}
 =\frac{195-154\delta}{507-176\delta},\qquad
D_2=D_3=\frac{\Delta+r}{b+2r}
 =\frac{156-11\delta}{507-176\delta}.
\end{equation*}
The denominators and components are positive throughout $[0,1]$, so this is the
target-1 conditional optimum on both open regimes.

It remains to compare the three conditional defender values.  For
$0\le\delta<\delta^*$, direct simplification gives
\begin{equation*}
V_3-V_1=-\frac{38(10\delta+39)}{176\delta-507}>0,
\qquad
V_3-V_2=\frac{836\delta-1911}{176\delta-507}>0.
\end{equation*}
For $\delta^*<\delta\le1$,
\begin{equation*}
V_3-V_2=\frac{341\delta-1677}{11(\delta-39)}>0,
\end{equation*}
while
\begin{equation*}
V_3-V_1=
\frac{59290\delta^2-211783\delta-388362}{286(176\delta-507)}>0.
\end{equation*}
For the last inequality, the numerator is negative on $[\delta^*,1]$ because
its derivative is negative there and its value at $\delta^*$ is negative; the
denominator is also negative.  Hence target 3 is globally optimal on both open
regimes.  Substitution gives \eqref{eq:values_left} and
\eqref{eq:values_right}.  Proposition~\ref{prop:threshold} handles the singular
point.  Taking one-sided limits yields $V_{\B}^{2}\to23/11$ and
$V_{\R}^{2}\to16/55$, proving value continuity; the policy limits are
$\langle47/163,58/163,58/163\rangle$ from the left and
$\langle0,58/163,58/163\rangle$ from the right.
\end{proof}

At $\delta=0$, \eqref{eq:policy_left}--\eqref{eq:values_left} recover the
terminal full-state solution
$\langle5/13,4/13,4/13\rangle$, $V_{\B}=15/13$, and $V_{\R}=8/13$, as required
when continuation values vanish.

\begin{remark}[Set-valued behaviour at the threshold]
\label{rem:uhc}
The one-sided policy limits are the vertices
$\langle47/163,58/163,58/163\rangle$ and
$\langle0,58/163,58/163\rangle$ of \eqref{eq:threshold_set}; hence every limit
of nearby optimal policies is threshold-optimal. The correspondence is therefore
upper hemicontinuous. It is not lower hemicontinuous, because the third vertex
$\langle0,105/163,58/163\rangle$ cannot be approached by the unique optima on
either side. The selected policy jumps when the resource-minimal rule switches
to $D_1=0$ at the singular node.
\end{remark}

\begin{figure}[H]
  \centering
  \includegraphics[width=\linewidth]{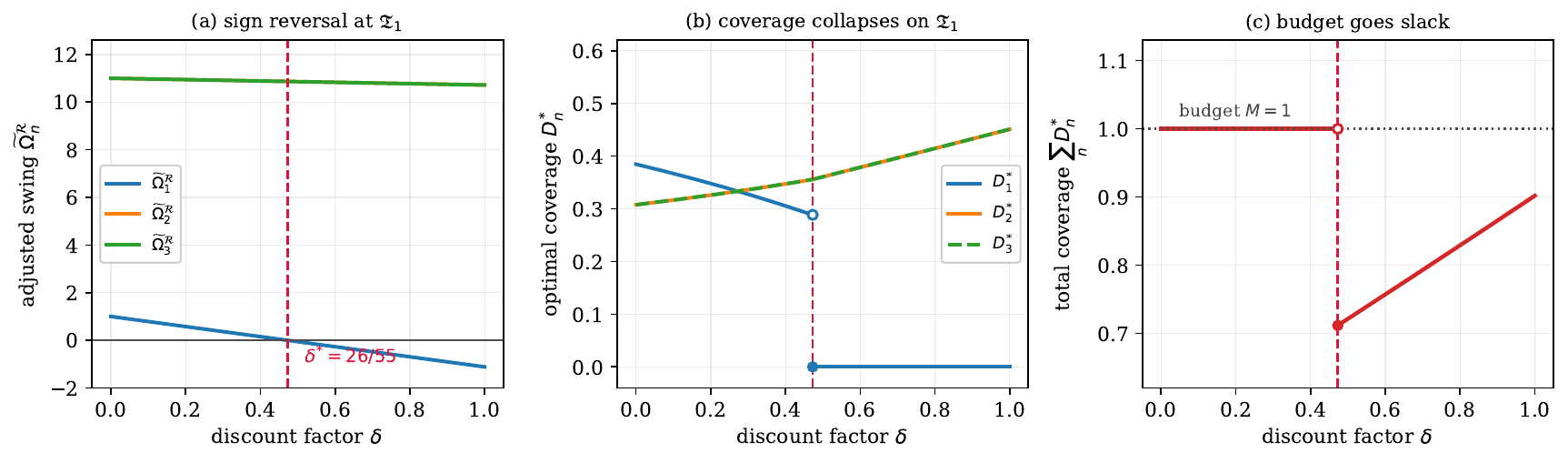}
  \caption{First-period solution of Example~\ref{ex:four} as the discount factor
  varies, with $\delta^{*}=26/55$ marked. (a) Adjusted attacker swings; only
  $\widetilde{\Omega}_{1}^{\R}$ changes sign. (b) Selected coverage under
  \eqref{eq:lexSSE}; the filled threshold point is the exact resource-minimal
  solution \eqref{eq:threshold_D}. (c) Total coverage against the budget $M=1$.
  The curves are the exact piecewise formulae of
  Theorem~\ref{thm:discount_global}.}
  \label{fig:sweep}
\end{figure}

Figure~\ref{fig:sweep} plots the exact formulas, which are reproduced by the
accompanying symbolic and rational-vertex checks. Coverage binds below the
threshold, drops to $116/163$ at the selected singular solution, and remains
slack above it.

\FloatBarrier

\section{Discussion and concluding remarks}
\label{sec:discussion}

The article's main contribution is the exact continuation adjustment for a
finite-horizon feedback Stackelberg attrition game. It preserves the static
affine target-payoff form at every node, allowing feasible full-support nodes to
be solved through one scalar attacker value and an endpoint comparison. The
reduction is deliberately local: boundary supports still require conditional
linear programmes, and the subset state space remains exponential.

More importantly, the adjustment changes the economics of coverage. An asset
may be attractive to preserve for future attacks, so covering it can increase
its current attack value. The worked instance derives this sign reversal from
strictly positive primitive swings and solves the global equilibrium for every
discount factor. At the exact singular threshold, the optimal set expands,
values remain continuous, and the selected policy discontinuously abandons the
asset favoured by the myopic allocation while releasing budget. This provides a
closed-form benchmark for dynamic security-game algorithms and a precise example
of how protection, deterrence, and resource use can diverge under attrition.

The most consequential restriction is the one recorded in
Remark~\ref{rem:outside}: the attacker is committed to strike in every period,
and with an outside option available the reversal effect vanishes in
Example~\ref{ex:four}. Establishing when reversal survives an outside option is
the natural next question, and would decide how far beyond forced-engagement
settings the mechanism reaches.

The results are further limited to complete information, perfect observation, identical
resources without routing constraints, and stagewise recommitment. Extensions
could seek structural conditions for full support at all reachable nodes,
characterise boundary supports analytically, or replace the uniform marginal
polytope with operationally constrained patrol sets.

\section*{Statements and declarations}

\noindent\textbf{Funding.} D.A. acknowledges support from the Australian
Research Council through Australian Laureate Fellowship FL240100217.

\noindent\textbf{Competing interests.} The authors declare no competing
interests.

\noindent\textbf{Author contributions.} All authors contributed to the
formulation of the model, the analysis and the writing of the manuscript, and
approved the final version.

\noindent\textbf{Data and code availability.} No empirical datasets were
generated or analysed. All numerical parameters are reported in the tables and
text. Supplementary Python scripts reproduce the budget calculations, the dynamic
recursion, the singular-threshold solution, the comparisons against
target-conditional programmes, and the figure.


\begin{thebibliography}{99}
\raggedright

\bibitem[Ba\c{s}ar and Olsder(1999)]{BasarOlsder}
Ba\c{s}ar T, Olsder GJ (1999) \textit{Dynamic Noncooperative Game Theory}, 2nd
edn. SIAM, Philadelphia. \url{https://doi.org/10.1137/1.9781611971132}

\bibitem[Ba\c{s}ar and Zaccour(2018)]{BasarZaccour}
Ba\c{s}ar T, Zaccour G (eds) (2018) \textit{Handbook of Dynamic Game Theory}.
Springer, Cham. \url{https://doi.org/10.1007/978-3-319-44374-4}

\bibitem[Bressan and Jiang(2020)]{BressanJiang2020}
Bressan A, Jiang Y (2020) Self-consistent feedback Stackel\-berg equilibria for
infinite-horizon stochastic games. Dyn Games Appl 10:328--360.
\url{https://doi.org/10.1007/s13235-019-00329-9}

\bibitem[Brown et~al.(2006)]{Brown2006}
Brown GG, Carlyle WM, Salmer\'on J, Wood K (2006) Defending critical
infrastructure. Interfaces 36(6):530--544.
\url{https://doi.org/10.1287/inte.1060.0252}

\bibitem[Cheng et~al.(2023)]{Cheng2023}
Cheng Z, Chen G, Hong Y (2023) Zero-determinant strategy in stochastic
Stackelberg asymmetric security game. Sci Rep 13:11308.
\url{https://doi.org/10.1038/s41598-023-38460-8}

\bibitem[Conitzer and Sandholm(2006)]{ConitzerSandholm2006}
Conitzer V, Sandholm T (2006) Computing the optimal strategy to commit to. In:
Proceedings of the Seventh ACM Conference on Electronic Commerce, pp 82--90.
\url{https://doi.org/10.1145/1134707.1134717}

\bibitem[Fudenberg and Tirole(1991)]{FudenbergTirole}
Fudenberg D, Tirole J (1991) \textit{Game Theory}. MIT Press, Cambridge, MA

\bibitem[Hohzaki and Nagashima(2009)]{Hohzaki2009}
Hohzaki R, Nagashima S (2009) A Stackelberg equilibrium for a missile
procurement problem. Eur J Oper Res 193(1):238--249.
\url{https://doi.org/10.1016/j.ejor.2007.10.033}

\bibitem[Hunt and Zhuang(2024)]{Hunt2024}
Hunt K, Zhuang J (2024) A review of attacker--defender games: current state and
paths forward. Eur J Oper Res 313(2):401--417.
\url{https://doi.org/10.1016/j.ejor.2023.04.009}

\bibitem[Iqbal et~al.(2024)]{Iqbal2024}
Iqbal A, Honhaga I, Teffera E, Perry A, Baker R, Pearce G, Szabo C (2024)
Vulnerability and defence: a case for Stackelberg game dynamics. Games 15(5):32.
\url{https://doi.org/10.3390/g15050032}

\bibitem[Jain et~al.(2010)]{Jain2010}
Jain M, Tsai J, Pita J, Kiekintveld C, Rathi S, Ord\'o\~nez F, Tambe M (2010)
Software assistants for randomized patrol planning for the LAX airport police
and the Federal Air Marshal Service. Interfaces 40(4):267--290.
\url{https://doi.org/10.1287/inte.1100.0505}

\bibitem[Kar et~al.(2018)]{Kar2018}
Kar D, Nguyen TH, Fang F, Brown M, Sinha A, Tambe M, Jiang AX (2018) Trends and
applications in Stackelberg security games. In: Ba\c{s}ar T, Zaccour G (eds)
\textit{Handbook of Dynamic Game Theory}. Springer, Cham, pp 1223--1269.
\url{https://doi.org/10.1007/978-3-319-44374-4_27}

\bibitem[Kiekintveld et~al.(2009)]{Kiekintveld2009}
Kiekintveld C, Jain M, Tsai J, Pita J, Ord\'o\~nez F, Tambe M (2009) Computing
optimal randomized resource allocations for massive security games. In:
Proceedings of AAMAS 2009, pp 689--696

\bibitem[Korzhyk et~al.(2010)]{Korzhyk2010}
Korzhyk D, Conitzer V, Parr R (2010) Complexity of computing optimal Stackelberg
strategies in security resource allocation games. In: Proceedings of AAAI-10,
pp 805--810. \url{https://doi.org/10.1609/aaai.v24i1.7638}

\bibitem[Korzhyk et~al.(2011)]{Korzhyk2011}
Korzhyk D, Yin Z, Kiekintveld C, Conitzer V, Tambe M (2011) Stackelberg vs Nash
in security games: an extended investigation of interchangeability,
equivalence, and uniqueness. J Artif Intell Res 41:297--327.
\url{https://doi.org/10.1613/jair.3269}

\bibitem[Kudo and Akimoto(2025)]{Kudo2025}
Kudo M, Akimoto Y (2025) Policy iteration for two-player general-sum stochastic
Stackelberg games. Proc Mach Learn Res 304:718--733.
\url{https://proceedings.mlr.press/v304/kudo26a.html}

\bibitem[Lauffer et~al.(2024)]{Lauffer2024}
Lauffer N, Ghasemi A, Hashemi H, Aggarwal V, Dullerud GE, Topcu U (2024)
No-regret learning in dynamic Stackelberg games. IEEE Trans Autom Control
69(3):1418--1431. \url{https://doi.org/10.1109/TAC.2023.3330797}

\bibitem[Li et~al.(2020)]{LiShenZheng2020}
Li H, Shen W, Zheng Z (2020) Spatial-temporal moving target defense: a Markov
Stackelberg game model. In: Proceedings of AAMAS 2020, pp 717--725

\bibitem[Li et~al.(2024)]{LiSojoudi2024}
Li J, Sojoudi S, Tomlin CJ, Fridovich-Keil D (2024) The computation of
approximate feedback Stackelberg equilibria in multiplayer nonlinear constrained
dynamic games. SIAM J Optim 34(4):3723--3749.
\url{https://doi.org/10.1137/24M1634709}

\bibitem[Osborne(2003)]{Osborne}
Osborne MJ (2003) \textit{An Introduction to Game Theory}. Oxford University
Press, Oxford

\bibitem[Pawlick et~al.(2019)]{Pawlick2019}
Pawlick J, Nguyen TTH, Colbert E, Zhu Q (2019) Optimal timing in dynamic and
robust attacker engagement during advanced persistent threats. In: Proceedings
of WiOpt 2019. \url{https://doi.org/10.23919/WiOPT47501.2019.9144123}

\bibitem[Pita et~al.(2008)]{Pita2008}
Pita J, Jain M, Marecki J, Ord\'o\~nez F, Portway C, Tambe M, Western C,
Paruchuri P, Kraus S (2008) Deployed ARMOR protection: the application of a
game-theoretic model for security at the Los Angeles International Airport. In:
Proceedings of AAMAS 2008: Industrial Track, pp 125--132

\bibitem[Shieh et~al.(2012)]{Shieh2012}
Shieh E, An B, Yang R, Tambe M, Baldwin C, DiRenzo J, Maule B, Meyer G (2012)
PROTECT: a deployed game theoretic system to protect the ports of the United
States. In: Proceedings of AAMAS 2012, pp 13--20

\bibitem[Sinha et~al.(2018)]{Sinha2018}
Sinha A, Fang F, An B, Kiekintveld C, Tambe M (2018) Stackelberg security games:
looking beyond a decade of success. In: Proceedings of IJCAI-18, pp 5494--5501.
\url{https://doi.org/10.24963/ijcai.2018/775}

\bibitem[Smith and Lim(2008)]{SmithLim2008}
Smith JC, Lim C (2008) Algorithms for network interdiction and fortification
games. In: Chinchuluun A et al (eds) \textit{Pareto Optimality, Game Theory and
Equilibria}. Springer, New York, pp 609--644.
\url{https://doi.org/10.1007/978-0-387-77247-9_24}

\bibitem[Smith and Song(2020)]{SmithSong2020}
Smith JC, Song Y (2020) A survey of network interdiction models and algorithms.
Eur J Oper Res 283(3):797--811.
\url{https://doi.org/10.1016/j.ejor.2019.06.024}

\bibitem[Tambe(2011)]{Tambe2011}
Tambe M (2011) \textit{Security and Game Theory: Algorithms, Deployed Systems,
Lessons Learned}. Cambridge University Press, Cambridge.
\url{https://doi.org/10.1017/CBO9780511973031}

\bibitem[Wu and Zhang(2026)]{WuZhang2026}
Wu Y, Zhang J (2026) Discounted stochastic Stackelberg games for peer-to-peer
energy sharing. Dyn Games Appl. \url{https://doi.org/10.1007/s13235-026-00703-4}

\bibitem[Yang et~al.(2014)]{Yang2014}
Yang R, Ford B, Tambe M, Lemieux A (2014) Adaptive resource allocation for
wildlife protection against illegal poachers. In: Proceedings of AAMAS 2014,
pp 453--460

\end{thebibliography}
\end{document}